%% file: z.tex
\documentclass[runningheads,11pt]{llncs}
\usepackage{graphicx}
\usepackage{fullpage}
\usepackage{color}
\usepackage{comment}
\usepackage{amsmath}
\usepackage{url}
\usepackage{multirow}
\usepackage{rotating}

\def\rank{\mbox{\rm {\sf rank}}}
\def\select{\mbox{\rm {\sf select}}}
\def\access{\mbox{\rm {\sf access}}}

\def\O{\mbox{\rm O}}
\DeclareMathSymbol{\Theta}{\mathalpha}{operators}{2}
\DeclareMathSymbol{\Phi}{\mathalpha}{operators}{8}

\def\+{\!+\!}
\def\-{\!-\!}

\begin{document}
\title{Rank, select and access in grammar-compressed strings\thanks{This research is partially supported by Academy of Finland grants 258308 and 250345 (CoECGR), JSPS grant KAKENHI 24700140, and by the JST PRESTO program.}}

\author{
Djamal Belazzougui\inst{1}
\and
Simon J. Puglisi\inst{1}
\and
Yasuo Tabei\inst{2}
}

\institute{
    Department of Computer Science,\\
    University of Helsinki\\
    Helsinki, Finland\\
    \email{\{belazzou,puglisi\}@cs.helsinki.fi}\\[1ex]
\and
    PRESTO,\\
    Japan Science and Technology Agency\\
    Saitama, Japan\\
    \email{tabei.y.aa@m.titech.ac.jp}\\[1ex]
}

\date{}

\maketitle \thispagestyle{empty}
\setcounter{page}{0}

\begin{abstract}
Given a string $S$ of length $N$ on a fixed alphabet of $\sigma$ symbols, a grammar
compressor produces a context-free grammar $G$ of size $n$ that generates $S$ and
only $S$. In this paper we describe data structures
to support the following operations on a grammar-compressed string: $\rank_c(S,i)$ (return
the number of occurrences of symbol $c$ before position $i$ in $S$); $\select_c(S,i)$ (return the
position of the $i$th occurrence of $c$ in $S$); and $\access(S,i,j)$ (return substring $S[i,j]$).
For $\rank$ and $\select$ we describe data structures of size $\O(n\sigma\log N)$ bits that support
the two operations in $\O(\log N)$ time. We propose another structure that uses
$\O(n\sigma\log (N/n)(\log N)^{1+\epsilon})$ bits and
that supports the two queries in $\O(\log N/\log\log N)$,
where $\epsilon>0$ is an arbitrary constant.
To our knowledge, we are the first to study the asymptotic complexity of $\rank$ and $\select$ in the grammar-compressed setting, and we provide a hardness result showing that significantly improving the bounds we achieve would imply a major breakthrough on a hard graph-theoretical problem.
Our main result for $\access$ is a method that requires $\O(n\log N)$ bits of space
and $\O(\log N+m/\log_\sigma N)$ time to extract $m=j-i+1$ consecutive symbols from $S$.
Alternatively, we can achieve $\O(\log N/\log\log N+m/\log_\sigma N)$ query time using
$\O(n\log (N/n)(\log N)^{1+\epsilon})$ bits of space.
This matches a lower bound
stated by Verbin and Yu for strings where $N$ is polynomially related to $n$.
\end{abstract}

\newpage
\section{Introduction} 
\label{sec:intro} 

Modern information systems and scientific experiments produce enormous amounts of 
highly repetitive data of various flavours. A particularly compelling example 
comes from the field of genomics, where recent breakthroughs in the biochemistry of DNA
sequencing have drastically reduced the cost and time required to produce whole 
genome sequences. This has in turn given rise to databases of tens of thousands of 
large but highly similar individual genome sequences~\cite{1000genomes,MNSV09jcb}.

Metagenomic samples~\cite{rsh2004}, increasingly used to describe the spicies diversity 
and makeup in microbial environments, are another kind of highly repetitive biological data. 
Each sequence in the dataset (called a read) is very short (say 200 characters),
but has high overlap (perhaps 180 characters, or more) with several other sequences in the
collection. A typical metagnomic sample is about a terabyte in size, but clearly also 
contains a great amount of redundancy.

Elsewhere, highly repetitive data takes the form of versioned document collections,
like source code repositories and other large multi-author collections, such as Wikipedia.
Even the web itself contains tremendous redundancy~\cite{FM2010} in the form of copied
and reused text and images~\cite{mbcmz2005,bc2009} and boilerplate. An extreme example
is the Web Archive\footnote{\url{http://archive.org/web/}} (or Wayback Machine), which 
contains regular crawls of the web (over 400 billion web pages in all) captured over the last 20 years.

In order to store the above types of data, compression must be employed. However,
effective compression alone is not enough: support for efficient queries over 
the original data without prior decompression is highly desirable.

{\em Grammar compression}~\cite{CLLPPSS2005} is particularly effective at compressing 
highly repetitive text data. 
Given a string $S$ of length $N$, a grammar compressor produces a context-free grammar $G$ 
that generates $S$ and only $S$. 
The size of the grammar refers to the total length of the right-hand sides of all rules. 
It is well known (see, e.g.,~\cite{n2012}) that on the kind of highly repetitive data mentioned above, 
grammar compressors (and their close relative LZ77~\cite{ZL77}) 
can achieve compression significantly better than statistical compressors, 
whose performance is expressed in terms of the popular $k^{\mbox{{\scriptsize th}}}$-order 
empirical entropy measure~\cite{M2001}. 

In this paper we consider support for three basic operations on grammar-compressed strings: 
\begin{align*}
  \access(S,i,j) &= \text{return the symbols in $S$ between $i$ and $j$ inclusive;} \\
  \rank_c(S,i) &= \text{number of occurrences of symbol $c\in \Sigma$ among the first $i$ symbols in $S$;} \\
  \select_c(S,j) &= \text{position in $S$ of the $j$th occurrence of symbol $c\in \Sigma $.}
\end{align*}

The $\access$ operation (often called ``random access'') allows one to process areas 
of interest in the compressed string without full decompression of all the symbols prior 
to $S[i,j]$. This is important, for example, in index-directed approximate pattern matching, 
where an index data structure first finds ``seed'' sites at which approximate matches to 
a query string may occur before more expensive alignment of the pattern by examining 
the text around these sites (via the $\access$ operation).

In their seminal paper Bille, Landau, Raman, Sadakane, 
Rao, and Weinmann~\cite{BLRSSW2011} show how, given a grammar of size $n$, it is possible 
to build a data structure of size $\O(n\log N)$ bits that supports access to any substring 
$S[i,j]$ in time $\O(\log N + (j-i))$. 

Operations $\rank$ and $\select$ are of great importance on regular (uncompressed) strings, where 
they serve as building blocks for fast pattern matching indexes~\cite{NM2007}, wavelet trees~\cite{GGV2003,N2014a}, 
and document retrieval methods~\cite{HPSTV2013,N2014b,NPV2011}. On binary strings, efficient rank 
and select support has been the germ for the now busy field of succinct data structures~\cite{M1996}.
Although many space-efficient data structures supporting rank and select operations 
have been presented~\cite{GMR2006,GGV2003,ggv2004,os2007,rrr2007}, they are 
not able to compress $S$ beyond its statistical entropy. 

To our knowledge we are the first to formally study algorithms for and the complexity of
rank and select in grammar-compressed strings, however three related results exist in the
literature.
Navarro, Puglisi and Valenzuela~\cite{NPV2011}, and more recently Navarro and Ord{\'o}{\~n}ez~\cite{Navarro14}, 
investigated practical methods for the rank operation in the context of indexed pattern matching. 
Their results pertain to grammars produced by a specific grammar compressor 
(the widely used RePair scheme~\cite{ml2000}), which are balanced and in Chomsky normal
form (i.e. straight-line programs~\cite{SLP}). They provide no formal analysis of the size of 
their data structure, but because their grammars are balanced their rank 
algorithm takes $O(\log{N})$ time. Experiments indicate the approach is practical for some applications. 

Recently, Bille, Cording, and G{\o}rtz~\cite{BCG2014}, used a weak form of select query, called 
{\em select-next}\footnote{These queries are referred to 
as {\em labeled successor queries} in~\cite{BCG2014}.} 
as part of their compressed subsequence matching algorithm. A select-next($S$,$i$,$c$) 
query returns the smallest $j > i$ such that $S[j] = c$. 

\paragraph{Our Contribution.} This paper provides the following results (also summarized in Table~\ref{tab-results}):
\begin{enumerate}
\item We show how to support $\access$ to $m$ consecutive symbols in $\O(\log N + m/\log_\sigma N)$ time 
using $\O(n\log N)$ bits of space. This is an improvement over the $\O(\log N + m)$ time solution of 
Bille et al., within the same space. Our scheme can be seen as a grammar-compressed counterpart to 
the 
scheme of Ferragina and Venturini~\cite{FV2007}
(see also Sadakane and Grossi~\cite{sg2006}), which has the same access time as ours, but only 
achieves $k^{\mbox{{\scriptsize th}}}$ order entropy compression (for some $k=o(\log_\sigma N)$).
\item We then show that by balancing the grammar and increasing space usage slightly, we obtain 
a data structure that supports $\access$ to $m$ consecutive symbols in $\O(\log N/\log\log N + m/\log_\sigma N)$ 
time and $\O(n\log N^{1+\epsilon})$ bits of space, matching a lower bound 
by Verbin and Yu~\cite{VY2013}.
\item We describe a data structure supporting $\rank$ and $\select$ operations in $\O(\log N)$ time 
and $\O(n\sigma\log N)$ bits of space for general (unbalanced) grammars, and $\O(\log N/\log\log N)$ 
time and $\O(n\sigma\log N^{1+\epsilon})$ bits of space for balanced grammars.
\item The above schemes for $\rank$ and $\select$ are fairly straightforward augmentations to
our $\access$ data structures, but our final result suggests that it is probably difficult 
to do much better. In particular, we show a reduction between $\rank$ and $\select$ operations in grammar
compressed strings and the problem of counting the number of distinct paths between two nodes in 
a directed acyclic graph, an old and seemingly difficult problem in graph theory. No significant progress 
has been made so far, even on the seemingly easier problem of reachability queries~\cite{cohen2003reachability}
(just returning whether the number of paths is non-zero). 
\end{enumerate}

\begin{table}[t]
\begin{center}
\caption{Summary of results for $\rank$, $\select$, and $\access$ in grammar-compressed strings.}
\begin{tabular}{|l|l|c|c|c|c|}
\hline
    &  & \multicolumn{2}{c|}{{\bf Rank/select}}   & \multicolumn{2}{c|}{{\bf Access}} \\
\hline
    &  & Unbalanced            & Balanced         & Unbalanced  & Balanced \\
\hline
\multirow{2}{*}{\bf This study} & Time  & $O(\log{N})$        & $O(\log{N}/\log\log{N})$  & $O(\log{N}+m/\log_{\sigma}{N})$ & $O(\log{N}/\log\log{N} + m/\log_{\sigma}{N})$ \\
\cline{2-6}
& Space & $O(n\sigma\log{N})$ & $O(n\sigma\log^{1+\epsilon}{N})$ & $O(n\log{N})$ & $O(n\log^{1+\epsilon}{N})$ \\
\hline \hline
\multirow{2}{*}{\bf Bille~\cite{BLRSSW2011}} & Time & $-$ & $-$ & $O(\log{N}+m)$ & $-$\\
\cline{2-6}
                                                     & Space & $-$ & $-$ & $O(n\log{N})$ & $-$\\
\hline \hline
\multirow{2}{*}{\bf Folklore} & Time & $-$ & $-$ & $-$ & $O(\log{N}+m)$ \\
\cline{2-6}
                                                     & Space & $-$ & $-$ & $-$ & $O(n\log N)$\\
\hline
\end{tabular}
\label{tab-results}
\end{center}
\end{table}

\paragraph{Related Work.} 

Apart from the above mentioned previous results for $\access$, $\rank$, and $\select$,  
there have been many results recently on 
grammar-compressed data structures for pattern matching~\cite{CN2010,CN2012,GGKNP2012},
document retrieval~\cite{NPV2011,NPV2013,NV2012}, and 
dictionary matching~\cite{INIBT2013b}.

There has also been a significant amount of recent research effort on developing 
string processing algorithms that operate directly on grammar-compressed data --- i.e., 
without prior decompression. To date 
this work includes algorithms for computing subword complexity~\cite{BCG2013,GBIT2013},
online subsequence matching and approximate matching~\cite{BCG2014,T2011}, faster edit 
distance computation~\cite{G2012,HLLW2013}, and computation of various kinds of biologically 
relevant repetitions~\cite{BGIILL2012,IB2012,IMSIBTNS2013,INIBT2013a}.

\section{Notation and preliminaries}
We consider a string $S$ of total length $N$ over an integer alphabet $[1..\sigma]$. 
The string $S$ is generated by a grammar that contains exactly $n$ non-terminals
(variables) and $\sigma$ terminals (corresponding to each character 
in the alphabet). 
We assume that the grammar is in Chomsky normal form (CNF). 
That is, each grammar rule is either of the form $R_0=R_1R_2$, where $R_1$ and $R_2$
are non-terminals, or of the form $R_0=c$, where $c$ is a terminal. The grammar 
has thus exactly $\sigma$ terminals. We note by $|R|$ the length of the string 
generated by the non-terminal $R$. 
In what follows, we only consider grammars in CNF, 
since every grammar of size $n$ with $\sigma$ terminals, 
can be transformed into a grammar in CNF of size $O(n)$
with $\sigma$ terminals. 

We will often use a DAG (directed acyclic graph) representation of the 
grammar with one source  (corresponding to the unique non-terminal that generates the whole string)
and $\sigma$ sinks (corresponding to the terminals). 
The height of a grammar is the maximal distance between the source 
and one of the sinks. The grammar is said to be balanced 
if its height is $O(\log N)$. Further, it is said to be 
weight-balanced if there exists some absolute constants 
$c_1$ and $c_2$ such that for any given rule $R_0=R_1R_2$, 
we have that $c_1|R_2|\leq |R_1|\leq c_2|R_2|$.

All our algorithms are in RAM model with word length at least $\log N$ bits, and 
with the assumptions that all the usual arithmetic and logic operations take 
constant time each. 
\section{Improved access time with rank and select support}

We now extend Bille et al.'s access scheme~\cite{BLRSSW2011} so that it uses the same space, 
$O(n\log N)$ bits (for a grammar of length $n$ that is not necessarily 
the smallest possible), but allows access to $m$ consecutive symbols in time $O(\log N+m/\log_\sigma N)$
instead of $O(\log N+m)$. We also show how to support rank and select 
within time $O(\log N)$ using $O(n\sigma\log N)$ bits of space. 
Before that, we start by reviewing the method of Bille et al. 

\subsection{The method of Bille et al.}
The method of Bille et al. uses heavy-weight paths in the DAG to support the access 
operation. More precisely, suppose the non-terminal that generates the whole string
is $R_0$. Given a position $x$, the method of Bille et al. generates 
a sequence of triplets $(R_1,s_1,e_1)$,$(R_2,s_2,e_2)$,...,$(R_t,s_t,e_t)$, where $t\leq \log N$, 
$(\sum_{1\leq i\leq t}{s_i-1})+1=x$ and $R_t$ is a non-terminal 
that generates a single character $c$. Note that we have $t \leq \log N$ because $|R_{i+1}|\leq |R_i|/2$
and $R_0\leq N$.
Thus $c$ is the character that is at position 
$x$ in the substring generated by the $R_0$.

Each triplet $(R_i,s_i,e_i)$ indicates that a substring generated by the non-terminal 
$R_i$ starts at position $s_i$ and ends at position $e_i$ inside the substring generated by the non-terminal $R_{i-1}$. The non-terminal $R_i$ is found using the heavy path 
that starts at the non-terminal $R_{i-1}$. This is explained 
in more detail in the next subsection. 

To every non-terminal $R_i$, there is an associated position $p_i$, which is 
the leaf in the heavy path that starts from the non-terminal $R_i$. We call $p_i$ the {\em center} 
of the non-terminal $R_i$. 

\subsection{Heavy path decomposition of a DAG}
Given a non-terminal $R$, the heavy path starting from the variable 
$R=P_0$ is the sequence of non-terminals $P_1,P_2,\ldots,P_t$, 
such that: 
\begin{enumerate}
\item For all $i\in[0,t-1]$, either $P_i=P_{i+1}Q_{i+1}$ or $P_i=Q_{i+1}P_{i+1}$ 
\item $|P_i|\geq |Q_i|$. 
\item The non-terminal $P_t$ generates a single character $c$. 
\end{enumerate}

Informally speaking, the heavy path starting from a non-terminal $P_0$
is the sequence of non-terminals $P_1,P_2,\ldots, P_t$ such that 
every non-terminal $P_{i+1}$ is the heaviest among the two non-terminals in the right-hand 
side of the non-terminal $P_i$ and the variable $P_t$ generates a single character. 
We associate with each non-terminal $R$ a center point $p$, obtained as follows: 
$$p=\sum_{1\leq i\leq k}|Q_{i_j}|+1$$
where the $i_j$ is the sequence of 
indices in $[1..t]$ such that $P_{i_j-1}=Q_{i+1}P_{i+1}$ (that is, $Q_{i_j}$ is the 
left non-terminal in the right-hand side of the non-terminal of $P_{i_j-1}$).  
The character at position $p$ in the string generated by $R$
is precisely the character generated by the non-terminal $P_t$. 

\begin{figure}
\begin{center}
\resizebox{0.4\textwidth}{!}{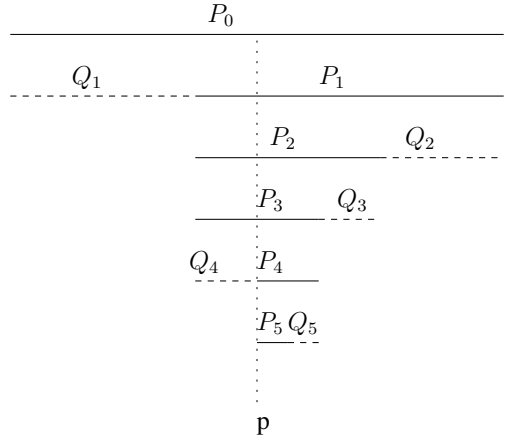}
\end{center}
\caption{An illustration of the heavy path starting from variable $P_0$ and its center point $p$.}
\label{fig-heavypath}
\end{figure}

\subsection{Biased Skip trees}

The main data structure used by Bille et al. is a forest
of trees built as follows. Each forest has as a root 
a node that corresponds to a sink in the DAG representation 
of the grammar. A node $x$ is a child of another node $y$ in the forest 
if and only if $y$ is the heavy child of $x$ in the original DAG. 
Thus the forest can be thought of as being built by reversing the original 
DAG (upside-down) and keeping only the edges that correspond to heavy paths. 
Then each of the resulting trees is represented using a 
biased skip tree. 
The representation allows us to find the sequence of triplets 
$(R_1,s_1,e_1),(R_2,s_2,e_2),\ldots,(R_t,s_t,e_t)$, given a position 
$x$ in the string $S$. 
Suppose that we are given a non-terminal $R$ with its center point $p$
and its heavy path $P_1, P_2, \ldots, P_t$. 
Given a point $x$ inside the substring generated by $R$
such that $x<p$, we wish to find inside the heavy path decomposition
of $R$ the non-terminal $P_i$ such that either: 
\begin{enumerate} 
\item $P_i=P_{i+1}Q_{i+1}$ with $x-p_i>|P_{i+1}|-p_{i+1}$, where $p_{i+1}$ is the center of the non-terminal $P_{i+1}$ and $p_i$ is the center of the non-terminal $P_i$. 
\item $P_i=Q_{i+1}P_{i+1}$ with $p_i-x>p_{i+1}-1$, where $p_{i+1}$ is the center 
the non-terminal $P_{i+1}$ and $p_i$ is the center of the non-terminal $P_i$.  
\end{enumerate}
Informally speaking, the non-terminal $P_i$ is the last non-terminal in the heavy path that contains the point 
$x$ and the non-terminal $Q_{i+1}$ hangs from the heavy-path, either from the right (first case above)
of from the left (second case above). 
The biased skip tree allows to find the non-terminal $P_i$ in time 
$O(\log(|R|/|Q_{i+1}|))$. 
Then, the algorithm produces the triplet $(R_i,s_i,e_i)$ by setting 
$R_i=Q_{i+1}$, $s_i=p-p_i+1$ and $e_i=s_i+|R_i|-1$ 
and starts the same procedure 
above by replacing $R$ by $R_1$, which results in the triplet $(R_2,s_2,e_2)$. 
The algorithm continues in this way until it gets the 
triplet $(R_t,s_t,e_t)$. The total running time of the procedure is 
$O(\log N)$, since $|R|=N$ and the successive running times $O(\log(|R|/|R_1|)), 
O(\log(|R_1|/|R_2|)),\ldots,O(\log(|R_{t-1}|/|R_t|))$ add up to $O(\log N)$ 
time by a telescoping argument. 
\subsection{Improved access time}
The above scheme can be extended to allow decompression
of an arbitrary substring that covers position $[x,x']$ 
in time $O(m+\log N)$, where $m=x'-x+1$ is the length of the 
decompressed substring. The decompression works as follows. 
We first find the sequence of triplets 
$(R_1,s_1,e_1),(R_2,s_2,e_2),\ldots,(R_t,s_t,e_t)$ 
corresponding to the point $x$. We then find the sequence 
of triplets $(R'_1,s'_1,e'_1),(R'_2,s'_2,e'_2),\ldots,(R'_t,s'_{t'},e'_{t'})$
corresponding to the point $x'$. 
We let $(R_i,s_i,e_i)=(R'_i,s'_i,e'_i)$ be the last common triplet between
the two sequences. Then the non-terminal $R_{i+1}$ hangs on the left 
of the heavy path that starts at the non-terminal $R_i$ 
and the non-terminal $R'_{i+1}$ hangs on the right. 
Without loss of generality, assume that $R_{i+1}$ hangs at a higher 
point than $R'_{i+1}$ and that $P_i$ is the last non-terminal on the heavy
path of $R_i$ that contains point $x$ (note that $P_i=R_{i+1}P_{i+1}$). 
Then the non-terminal $P_{i+1}$ still contains the point $x'$ and we need to decompress 
all the non-terminals that hang on the right of the heavy path that starts 
at $P_{i+1}$ down to the non-terminal from which the non-terminal $R'_{i+1}$ hangs. 
Afterwards, we just need to 1) decompress all the non-terminals that hang on the 
right of the heavy path that starts at $R_{j}$ down to the non-terminal from which 
$R_{j+1}$ hangs for all $j\in[i+1,t-1]$, and then 2) symmetrically 
decompress all the non-terminals that hang on the 
left of the heavy path that starts at $R'_{j}$ down to the non-terminal from which 
$R'_{j+1}$ hangs for all $j\in[i+1,t'-1]$. 
This whole procedure takes $O(m+\log N)$ time. The main point of the procedure 
is to be able to decompress all the non-terminals that hang 
on the right or on the left of the portion of some heavy path that starts at some 
non-terminal $P_i$ down to some non-terminal $P_{i+1}$ inside that heavy path. 

In what follows, we show how to reduce the time to just $O(m/\log_\sigma N+\log N)$. 
For every non-terminal $X$ that generates a right-hand side $YZ$, we will 
store the following additional fields, which occupy $O(\log N)$ bits: 
\begin{enumerate}
\item The $\log N/\log\sigma$ leftmost characters in the 
substring generated by $X$. 
\item The $\log N/\log\sigma$ rightmost characters in the 
substring generated by $X$. 
\item Three jump pointers. Each jump pointer is a pair
of the form $(R,p)$, where $R$ is non-terminal 
and $p$ is a position inside the substring generated by $X$. 
\end{enumerate}
The three jump pointers are called left, right and central
(any of them may be empty). 
The jump pointers will allow us to accelerate the extraction 
of characters. 
The central jump pointer  allows to 
fully decompress any given non-terminal that generates 
a string of length $m$ in time $O(m/\log_\sigma N)$.
The right and left jump pointers will allow us to jump along the heavy paths, 
avoiding the traversal of all the non-terminals in the heavy path. 
We first start by describing the central jump pointer (henceforth called central field). 
\subsubsection{Non-terminal decompression}
The central field of $X$ will contain a pointer to another variable
in addition to the position, where the substring of that variable 
starts inside the string generated by $X$. 
If $X$ generates a string of length less than $2\log_\sigma N$, 
then we do not store the central field at all. 
Otherwise we do the following. We use two temporary counters $C_\ell$ and $C_r$.
If $Y$ generates a string of length less than $\log_\sigma N$, 
then we increment $C_\ell$ by $|Y|$. If $Z$ generates a string of length less than $\log_\sigma N$, 
then we increment $C_r$ by $|Z|$. 
If one of the two variables $Y$ or $Z$ generates a string of length 
at least $\log_\sigma N$, we let $W$ be that variable and let 
its right-hand side be $UV$. We now show a recursive procedure 
implemented on the variable $W$. 
If both $U$ and $V$ are of length at least $\log_\sigma N$, we stop the recursion. 
If either of $U$ or $V$ is of length 
less than $\log_\sigma N$, we increment $C_\ell$ by $|U|$
if $C_\ell+|U|$ or increment $C_r$ by $|V|$ if 
$C_r\leq \log N/\log\sigma$. 
Otherwise, if we succeeded in incrementing $C_\ell$, we recurse on 
the variable $V$ by setting $W=V$. If we succeeded in incrementing 
$C_r$, we recurse on the variable $U$ by setting $W=U$. 
Whenever we stop for a variable $W$ and could no longer recurse, 
we set the central field to $(W,C_\ell+1)$. 

The decompression is done in a straightforward way. 
Whenever we want to decompress a variable that generates 
a string of length at most $2\log N/\log\sigma$, we just write the left and right substrings of lengths 
$\log N/\log\sigma$, which are already stored with the variable. 
Whevener we want to decompress a variable $X$ that has a central field $(R,p)$, 
we decompress everything inside the substring generated by $X$ that lies 
on the left and right of $R$ (that is, substrings that span characters $[1,p-1]$ 
and $[p+|R|,|X|]$) which are both stored inside the left 
and right substrings of $X$. We then recurse on the variable $R$. Otherwise, if 
$X$ does not have a central field simply recurse on the two variables 
of its right-hand side.
We will now prove that the total time to decompress any variable that generates 
a string of length $m$ is $O(1+m/\log_\sigma N)$. 

To prove this, we will analyze the virtual decompression tree 
traversed during the decompression and prove that the total number of 
nodes in such a tree is $O(1+m/\log_\sigma N)$. To that purpose 
we will first prove that the number of leaves is at most $O(1+m/\log_\sigma N)$. 
In order to reach a leaf $X$ during the decompression, there could be two cases. 
\begin{enumerate}
\item We reach $X$ from a node $Y$ that has two children of weight at least $\log_\sigma N$ 
and one of them is $X$ and we followed the pointer to $X$.
\item We reach $X$ from a node $Y$ that contains in its central field a pointer to $X$. 
\end{enumerate}
In the first case, $X$ generates a string of length at least $\log N/\log\sigma$. 
Thus, there can be at most $m/(\log_\sigma N)$ such leaves since they all generate 
disjoint substrings. 

If we reach $X$ from a node $Y$ that has a central field, then necessarily $Y$ generates 
a substring of length more than $2\log_\sigma N$. We can therefore uniquely associate 
$Y$ with $X$ and thus conclude that there can be at most $m/(2\log_\sigma N)$ such leaves. 
We have proved that the number of leaves is at most $O(m/\log_\sigma N)$. This implies 
that the number of nodes of degree two in the decompression tree is at most $O(m/\log_\sigma N)$. 
It remains to show that the number of unary nodes in the decompression tree is at most 
$O(m/\log_\sigma N)$. For that, it suffices to show that we decompress a substring 
of length $\Theta(\log_\sigma N)$ everytime we traverse two consecutive unary nodes $N_0$ and then $N_1$
in the decompression tree. This is obvious, since if we do not do so, then we can merge both the right 
substring and the left substring of $N_1$ into the ones of $N_0$ and delete $N_1$. 
We thus have proved the following lemma.
\begin{lemma}
Suppose that we are given a grammar of size $n$ that generates a string $S$ of length $N$ over an 
alphabet of size $\sigma$ is of size $n$. Then we can build a data structure that uses 
$O(n\log N)$ bits so that decompressing the string generated by any non-terminal 
takes $O(\log N+m/\log_\sigma N)$ time, where $m$ is the length of the string 
generated by the non-terminal. 
\end{lemma}

\subsubsection{Decompression of arbitrary substrings}
We now show how to decompress an arbitrary substring 
that is not necessarily aligned on a non-terminal. 

Recall that the core procedure for decompression is as follows. 
We are given a non-terminal $R$ and another non-terminal
$Q$ that hangs from the heavy path down from $R$ and we want to decompress 
what hangs from the left (respectively right) of the heavy path down 
to the point where $Q$ hangs from the heavy path. To accelerate the decompression 
we will use the right and left jump pointers. Since the right and left 
decompression are symmetric, we only describe the case of left decompression. 

Before describing the decompression itself, 
we first describe how the left jump pointer for non-terminal $R$ is set. 
We will keep a counter $C$, assume $P_0=R$ and inspect the sequence $P_1,P_2,\ldots,P_t$. For every 
$i$ starting from $1$ such that $P_{i-1}=Q_iP_i$, we increment $C$ 
by $|Q_i|$ and stop when $C+|Q_i|>\log N/\log\sigma$. Then the left jump pointer
will point to the non-terminal $P_{i-1}=L$ along with its starting point $p_L$ inside $R$. 
If $P_0=Q_1P_1$ and $|Q_1|>\log N/\log\sigma$ or $C+|Q_i|$ never exceeds 
$\log N/\log\sigma$, then we do not store a left jump pointer at all. 

The decompression of the left of a heavy path is done as follows. We are given the non-terminal $Q$
and its starting position $p_Q$ inside $P$. 
At first, we first check whether $p_Q\leq \log N/\log\sigma$, in which case 
everything to the left of $Q$ inside $P$ is already in the left substring of $P$
and it can be decompressed in time $O(1+p_Q/\log_\sigma N)$ and we are done. 

If $p_Q>\log N/\log\sigma$, then we have two cases:
\begin{enumerate}
\item If $P$ has a left jump pointer $(L,p_L)$ then $p_Q\geq p_L$
and we decompress the first $p_L\leq \log N/\log\sigma$ characters 
of the string generated by $P$ (from the left substring of $P$), then replace $P$ by 
$L$ and recurse on $L$ and $Q$. 
\item If $Q$ does not have a left jump pointer, then 
we necessarily have that $P=Q_1P_1$ with $|Q_1|>\log N/\log\sigma$
and $p_Q>|Q_1|$, we just decompress $Q_1$ (using the procedure shown above 
for fully decompressing non-terminals using central jump pointers). 
replace $P$ by $P_1$ and recurse on $P_1$ and $Q$. 
\end{enumerate}
It remains to show that the bound for the procedure is $O(1+y/\log_\sigma N)$,
where $y$ is the total length of the decompressed string. 
Analyzing the recursion, it can easily be seen that 
when we follow two successive left jump pointers, 
we are decompressing at least $\log_\sigma N$ characters from left substrings. 

Otherwise, if we do not follow a jump pointer, then we are either decompressing 
a non-terminal of length at least $\log_\sigma N$ characters in optimal time
or we terminate by decompressing at most $\log_\sigma N$ characters. 

We thus have shown the following theorem. 

\begin{theorem}
Suppose that we are given a grammar of size $n$ that generates a string $S$ of length $N$ over an 
alphabet of size $\sigma$ is of size $n$. Then we can build a data structure that uses 
$O(n\log N)$ bits that supports the access to $m$ consecutive 
characters in time $O(\log N+m/\log_\sigma N)$ time. 
\end{theorem}
\subsection{Rank and select}
In order to support rank we will reuse the same forest of biased skip trees used by Bille et al. 
(see above). For every node $\alpha$ representing 
a non-terminal $R$ with center point $p_R$, we associate 
$\sigma$ values $v_{R,1},\ldots v_{R,\sigma}$, where $v_{R,c}$ represents the number of occurrences 
of character $c$ in the interval $R[1,p_R]$. The rank operation can be supported by using the values $v_i$ stored at each node. 
That is counting the number of occurrences of character $c$ before position $p$ in the string 
generated by $R=R_0$ can be done by first finding the sequence of triplets (see above)
$(R_1,s_1,e_1)$,$(R_2,s_2,e_2)$,...,$(R_t,s_t,e_t)$, where $R_i$ is in the right hand-side 
of some non-terminal $P_{i}$ that is the heavy path of $R_{i-1}$. Then the of result of the rank 
operation will clearly be $$\sum_{i=0}^{t-1}(v_{R_i,\sigma}-v_{P_{i+1},\sigma})$$ 
The total space will be $O(n\sigma\log N)$ bits, while the query time still remains $O(\log N)$. 

In order to support select, we will construct $\sigma$ DAGs, where DAG number $c$ is built as follows. 
We remove every non-terminal whose corresponding substring does not contain 
the character $c$. We also remove every non-terminal for which one of the two 
non-terminals in the right-hand side generates a string that does not contain $c$. 
We then construct a biased skip tree rooted at the terminal $c$, but replacing the size of the 
strings generated by the non-terminals with the number of occurrences of character $c$ in the 
string generated by the non-terminal (the heavy paths are constructed by taking into account 
the number of occurrences of character $c$ instead of the total size of the strings generated 
by the non-terminals). 
Given a non-terminal $P$, the centerpoint $p_c$ determined by the heavy path will indicate 
the occurrence number $p_c$ of character $c$. We now associate the value $u_{P,c}$ which stores the 
position of the occurrence number $p_c$ inside the substring generated by $c$. 
Given a non-terminal $R=R_0$ and an occurrence number $x$ of $c$, we can determine 
the position $p$ of terminal $c$ in the substring generated by $R$ as follows: 
we first use the biased skip tree to determine the sequence 
$(R_1,s_1,e_1)$,$(R_2,s_2,e_2)$,...,$(R_t,s_t,e_t)$,
where $R_i$ is in the right hand-side of some non-terminal $P_{i}$ that is the heavy path of $R_{i-1}$.
Then the result of the select query will be 
$\sum_{0\leq i\leq t-1}(u_{R_i,\sigma}-u_{P_{i+1},\sigma})+1$. 
Overall the time taken to answer a select query will be $O(\log N_c)\leq O(\log N)$, where 
$N_c$ is the number of occurrences of character $c$ in the whole string. 
The total space will clearly be $O(n\sigma\log N)$ bits of space. 
We have proved the following. 
\begin{theorem}
Suppose that we are given a grammar of size $n$ that generates a string $S$ of length $N$ over an 
alphabet of size $\sigma$. Then we can build a data structure that uses 
$O(n\sigma\log N)$ bits that supports {\em rank} and {\em select} queries in time $O(\log N)$. 
\end{theorem}
\section{Optimal access time for not-so-compressible strings}
\begin{theorem}
Given a weight-{\em balanced} grammar of size $n$ generating a string $S$ of length $N$ over an
alphabet of size $\sigma$, we can build a data structure that uses
$O(n\log^{1+\epsilon} N)$ bits (for any constant $\epsilon$), that 
supports random access to any character of $S$ in $O(\log N/\log\log N)$ 
time, and access to $m$ consecutive characters in $O(\log N/\log\log N+m/\log_\sigma N)$ time.
Furthermore, we can build a data structure that uses $O(n\sigma\log^{1+\epsilon} N)$
bits of space and that supports rank and select in $O(\log N/\log\log N)$ time.
\end{theorem}

Because the grammar is balanced it will have the property that
any root-to-leaf path in the corresponding tree will have depth $O(\log N)$~\cite{R2003,CLLPPSS2005}.
For every variable $X$ we generate a new right-hand side that contains at most $\log^\epsilon N$
variables by iteratively expanding the right-hand side of $X$ down $\epsilon\log\log N$ levels (or less
if we reach a terminal). We then store in a fusion tree~\cite{FW90,FW93}
the prefix-sum of the lengths of the strings generated by each variable in the right-hand side.
Every fusion tree uses $O(\log^{1+\epsilon}N)$ bits of space and allows 
constant time predecessor search on the prefix-sums. More precisely, assuming that the expanded
right-hand side of $R$ is $R_1R_2\ldots R_t$ with $t\leq 2^{\epsilon\log\log n}$, 
the prefix sums are $v_1,v_2\ldots v_{t}$, where $v_i=|R_1|\ldots |R_{i-1}|$. 

The fusion tree is a data structure which uses linear space and allows predecessor searches 
on a set of $t$ integers of $w$ bits in $O(\log t/\log w)$. Since in our case, we have 
$t=\log^\epsilon N$ and $w\geq \log N$, the query time is constant. 
It is clear that a top-down traversal of the tree can now be done in time $O(\log N/\log\log N)$.
This is because we are moving $\epsilon\log\log N$ levels at a time at each step and a leaf
is reached after descending at most $O(\log N)$ levels.

We note that our upper bound for access matches the lower
bound of Verbin and Yu~\cite{VY2013} who have shown that for ``not-so-compressible'' strings
--- those that have a grammar of length $n$ such that $N\leq n^{1+\epsilon}$
for some constant $\epsilon$ --- the query time cannot be better
that $O(\log n/\log\log n)=O(\log N/\log\log N)$ if the used space 
is not more than $O(n\log^c n)$ for some constant $c$.
Extending the scheme to select and rank queries is not so hard and
multiplies the space by a factor $\sigma$.
For supporting rank queries it suffices to augment every node that corresponds 
to a variable $R$ with the prefix-sums 
of the number of occurrences of character $c$ in the strings generated 
by the variables in the expanded right-hand side for every character $c$. 
That is supposing that a variable $R$ has an expanded right-hand-side $R_1\ldots R_t$
with $t\leq 2^{\epsilon\log\log n}$, we store $t$ values $v_{1,c},\ldots v_{t,c}$ 
such that $v_{i,c}$ is the sum of the number of occurrences of character $c$ 
in the strings generated by rules $R_1,\ldots R_i$. 
Then a rank query can be answered during 
a top-down traversal of the DAG. We initially set a counter to zero. 
Then whenever we traverse a node corresponding to the variable 
$R$ with right-hand side $R_1,\ldots R_t$, and the fusion tree indicates 
to continue through the node $R_i$, we add the value $v_{i-1}$ to the counter 
(or zero if $i=1$) and at the end return the value of the counter. 
 
For select queries, we can build $\sigma$ DAGs, where in the DAG for character $c$ 
we replace the length of the strings 
by the count of character $c$. Answering select queries for character $c$ is then easily done by a top-down traversal
of the DAG with expanded right-hand side. Suppose that for a node corresponding 
to a variable $R$, we store the sequence $v_{1,c},v_{2,c},\ldots c_{i,c}$
in a fusion tree associated to variable $R$ and character $c$, 
where $v_{i,c}$ is the total number of occurrences of character $c$ 
in $R_1,\ldots R_i$. We start by setting a counter $C=0$ and then traverse the DAG top-down. 
For each traversed node, we query a fusion tree 
associated with character $c$ and a rule $R$ 
to determine which variable in the expanded right-hand side of $R$ to 
continue the traversal to. If the variable indicated is $R_i$, then 
we increment $C$ by $v_i=|R_1|+\ldots |R_{i-1}|$ (the value $v_i$ is stored 
in the original DAG used to support access queries). 

It remains to show how  access queries are supported. 
We can show that in time $O(\log N/\log\log N+m/\log_\sigma N)$
we can decompress a substring of length $m$ at an arbitrary position, while
using the same $O(n\log^{1+\epsilon}N)$ bits of space.

Consider any variable $X$ that generates two variables $Y$ and $Z$.
If $Y$ generates a string of length at most $\log_\sigma N$,
then we keep the whole string generated by $Y$ in the node
corresponding to $X$ instead of writing the identifier of $Y$.
Similarly if $Z$ generates a string of length at most $\log_\sigma N$,
we keep in the node $X$ the whole string generated by $Z$ instead
of keeping the identifier of $Z$.

We will now prove that decompressing the substring generated 
by any variable $X$ can be done in time $O(1+|X|/\log_\sigma N)$. 
The decompression is done as follows. If both $|Y|\leq \log_\sigma N$ 
and $|Z|\leq \log_\sigma N$, then we directly output the substrings 
of $X$ and $Y$, since the 
two strings are stored in $X$. If we have $|Y|\leq \log_\sigma N$ 
and $|Z|>\log_\sigma N$, then we output $Y$ and recurse on $Z$.
If we have $|Z|\leq \log_\sigma N$ 
and $|Y|>\log_\sigma N$, then we output $Z$ and recurse on $Y$. 
If we have both, i.e.,~$|Y|>\log_\sigma N$ 
and $|Z|>\log_\sigma N$, then we recurse on both $Y$ and $Z$. 

We will now analyze the decompression by proving that the (virtual) 
decompression tree contains always at most $O(1+|X|/\log_\sigma N)$ nodes. 
We assume that $|X|=m\geq \log_\sigma N$. Otherwise the decompression 
of $X$ is immediate since both $|Y|\leq \log_\sigma N$ 
and $|Z|\leq \log_\sigma N$. 
We first prove that the number of leaves is $O(m/\log_\sigma N)$, 
Since each leaf must contain a string of length at least $\log_\sigma N$ 
(recall that the string corresponding to variable of length less than $\log_\sigma N$
is stored in the corresponding parent of the leaf), we must output at least $\log_\sigma N$ 
characters per leaf and thus we cannot have more than $m/\log_\sigma N$ leaves. 
The number of nodes of degree at least $2$ in any tree
is less than the number of leaves, and those nodes correspond to the case 
where both $|Y|>\log_\sigma N$ 
and $|Z|>\log_\sigma N$. It remains to bound the number of unary internal nodes 
which correspond to the case where either $|Y|\leq \log_\sigma N$ and $|Z|>\log_\sigma N$
or $|Z|\leq \log_\sigma N$ and $|Y|>\log_\sigma N$. We note that in this case, 
we output the substring associated with either $Y$ or $Z$. Because the grammar 
is weight-balanced, and $|X|\geq \log_\sigma N$, it is easy to see that $Y\geq c_0\log_\sigma N$
and $Z\geq c_0\log_\sigma N$. We thus output a string of length $c_0\log_\sigma N$ 
every time we traverse a unary node. 

We now show how to decompress an arbitrary substring of length $m$ 
in time $O(\log N/\log\log N+m/\log_\sigma N)$. 
For that, we build the access structure above for each variable.
In addition, for each variable $X$, we store the first and the last 
$\log_\sigma N\log^\epsilon N$ characters of the substring generated by $X$. 
This adds $O(\log^{1+\epsilon} N)$ bits of space per variable. 
The decompression of a substring that spans positions 
$[x,y]$ inside the $R$ will first start by traversing the DAG 
top-down for position $x$ and $y$. We collect the two paths that consist 
of the $O(\log N/\log\log N)$ traversed nodes for positions $x$ and $y$. 
We note by $R$, the variable that corresponds to the lowest common 
node between the two paths. The variable $R$ will have an expanded right-hand-side 
$R_1R_2\ldots R_t$, such that the next nodes on the paths of positions 
$x$ and $y$ have respectively associated variables $R_i$ and $R_j$ with $j>i$. 
We thus need to decompress all variables $R_k$ for $i<k<j$ and then decompress 
the prefix of the string generated by $R_j$ up to the position that corresponds 
to $y$ and the suffix of the string generated by $R_i$ up to the position that corresponds 
to $x$. 
It remains to show how to decompress a prefix of the string generated 
by a variable $R$ up to an arbitrary position $x$ in optimal time. 
Suppose that $R=R_0R_1\ldots R_t$ and that $x$ is in variable 
$R_i$. We will have to decompress the substring generated by 
$R_0R_1\ldots R_{i-1}$. In order to do that we first check whether 
$\ell=|R_0|+|R_1|+\ldots |R_{i-1}|\geq \log^\epsilon N\log_\sigma N$. 
If that is the case, we decompress all the variables $R_0,R_1,,\ldots R_t$
in time $O(\ell/\log_\sigma N+i-1)$. 
Since $i-1\leq t\leq \log^\epsilon N$ and $\ell/\log_\sigma N\geq \log^\epsilon N$, 
we deduce that the time is $O(\ell/\log_\sigma N)$, which is optimal. 
If $\ell<\log^\epsilon N\log_\sigma N$, we can directly decompress the first 
$\ell$ characters of the substring generated by $R$ in optimal $O(1+\ell/\log_\sigma N)$ 
time, since we have stored the first $\log^\epsilon N\log_\sigma N$ characters of $R$. 
In the subsequent steps, we can continue from $R_i$ and the relative position 
of $x$ in the substring generated by $R_i$. Since we traverse at most 
$\log N/\log\log N$ levels and we spend optimal time for decompression at each 
level, the total time is $O(\log N/\log\log N+m/\log_\sigma N)$, which is optimal.

Decompressing the suffix generated 
by a variable $R$ up to an arbitrary $x$ is symmetric.

We keep the consecutive nodes that we have traversed in a stack. Once we reach a leaf (which is a node
generating a substring of length between $\log_\sigma N$ and $2\log_\sigma N-2$),
we can get the leaf's content in $O(1)$ time. We then traceback in the tree, collecting 
the substrings
contained in the nodes or the leaves until we will have decompressed $O(m)$
characters. It is easy to see that we will not traverse more than
$O(\log N/\log\log N+\log_\sigma N)$ nodes. 

\begin{corollary}
Given an arbitrary context-free grammar of size $n$ generating a string $S$ of length $N$ over an
alphabet of size $\sigma$, we can build a data structure that uses
$O(n\log (N/n)\log^{1+\epsilon}N)$ bits (for any constant $\epsilon$) such that 
access to $m$ consecutive
characters supported in time $O(\log N/\log\log N+m/\log_\sigma N)$ time.
Furthermore we can build a data structure that uses $O(n\sigma\log (N/n)\log^{1+\epsilon}N)$
bits of space and that supports rank and select in $O(\log N/\log\log N)$ time.

\end{corollary}
To achieve the corollary we can use the scheme of Charikar et al.~\cite{CLLPPSS2005}, 
which can generate a weight-balanced grammar from an unbalanced one such that the balanced 
grammar generates the same string as the unbalanced one, but is of size larger by a factor 
at most $O(\log(N/n))$.

\section{The hardness of rank and select in grammar-compressed strings}
\label{sec:lowerbound}

We will now show a reduction from the problem of rank and select in grammar compressed
strings to path counting in DAGs.

Suppose that we are given a DAG with $m$ nodes
and $n$ edges that has $\beta$ nodes with indegree $0$ (sources) and $\sigma$ nodes with outdegree
$0$ (sinks) and we want to answer the following type of query:
given any node $u$ and any sink $v$, compute the number of distinct paths that connect
$u$ to $v$. 
We allow multi-edges. That is, we allow 
more than one edge $uv$ with the same nodes $u$ and $v$.
Let $N$ be the total number of distinct paths that connect
the $\beta$ sources to the $\sigma$ sinks.  
We can show that we can construct a grammar-compressed string of (uncompressed) length
$N$ that contains $O(n)$ non-terminals and $\sigma$ terminals and such that
answering the described query on the graph reduces to answering rank queries
on the grammar compressed string.

We modifiy the DAG such that it contains a single source and
all nodes it contains are either of outdegree $0$ or $2$:
\begin{enumerate}
\item For every node $v$ of outdegree $1$ do the following. If the 
successor of the node is $w$, then for every edge $uv$ create 
a new edge $uw$ and remove the edge $uv$. After making these changes, 
node $v$ will become of indegree $0$. 
We remove the node $v$ and keep a data structure that maps the node $v$ to node $w$. 
Since any path starting at $v$ must pass through $w$, we will know that counting 
the number of paths starting at $v$ and ending at a node $x$ will be exactly the 
same as the number of paths starting at $w$ and ending in $x$. Note that all the paths 
that went through node $v$ will be preserved. Thus the count of the number of paths 
will not change. 
\item If the number of nodes having indegree $0$ is $t\geq 2$, then create
a new root and connect the root to all the nodes of indegree $0$ by creating
$t-2$ intermediate nodes. The root and the newly created nodes will have outdegree $2$.
\item For every node $v$ of outdegree $d\geq 3$, we will add exactly $d-2$
intermediate nodes of outdegree $2$ that connect the original nodes
with the destination and modify $v$ so that it has outdegree $2$.
\end{enumerate}
Clearly, the constructed DAG will have $O(m)$ nodes and will
generate a string of length exactly $N$, where $N$ is the total
number of distinct paths between one of the original $\beta$
sources and one of the original $\sigma$ sinks. Figure~\ref{fig-transform} in the appendix illustrates this transformation.

For every non-terminal, we will store two pointers that delimit the leftmost
occurrence of the rule in the text (the beginning and ending positions).
This array occupies $O(n)$ words of space.
Then, in time $T(n,\sigma,N)$, we build a data structure
of size $S(n,\sigma,N)$ that answers rank queries on the
string generated by the grammar in time $t(n,\sigma,N)$.
To answer a query that counts the number of paths between a node
$u$ and a designated sink $v$, we will
find the non-terminal $R$ that corresponds to $u$ and the terminal $c$ that corresponds
to $v$. We then find (using the array of pointers) the two positions $i$ and $j$ that correspond to the leftmost
occurrences of $R$ in the text. Finally, the count is returned by doing two rank queries
for symbol $c$ at positions $i$ and $j$ respectively.

\begin{theorem}
Suppose there exists a scheme that can preprocess a grammar of
size $n$ with $\sigma$ non-terminals that generates a
string of length $N$ in $T(n,\sigma,N)$ time and produces a data structure
of size $S(n,\sigma,N)$ that answers to rank queries on the
string generated by the grammar in time $t(n,\sigma,N)$. Then given a
DAG with $m$ nodes, $n$ edges (possibly with multiedges), 
$\beta$ sources and $\alpha$ sinks, we can, after preprocessing time $O(n+T(n,\sigma,N))$, 
produce a data structure of size $O(n+S(n,\sigma,N))$ that can count the number of
distinct paths from any node of the DAG to one of the $\sigma$ sinks
in time $O(t(n,\sigma,N))$, where $N$ is the number of distinct paths
that connect the $\beta$ sources to the $\alpha$ sinks.
\end{theorem}

\section{Concluding Remarks}
\label{sect:conlusion}

Perhaps the most important open question we raise is whether our results for $\rank$ and $\select$ 
are optimal. As we have shown, proving this one way or the other would lead to progress on
the path counting problem in DAGs, an old and interesting problem in graph theory. Relatedly,
similar approaches may prove fruitful in judging the hardness of other problems on grammar-compressed
strings, many solutions to which currently seem to be loose 
upperbounds~\cite{BCG2014,BGIILL2012,IB2012,IMSIBTNS2013,INIBT2013a}.

Our result for $\access$ closes the gap between Bille et al.'s random access result~\cite{BLRSSW2011} 
and the lowerbound of Verbin and Yu~\cite{VY2013} for the (large) set of strings whose 
grammar-compressed size $n$ is polynomially related to $N$. We leave closing the gap 
for the remaining strings as an open problem.
 
Another interesting problem for future work is to support these queries in compressed text whose size
is bounded in terms of the size of the LZ77 parsing, $z$, which is a lowerbound on $n$. The 
block graph data structure of Gagie, Gawrychowski and Puglisi~\cite{GGP2011} supports access queries 
for $m$ symbols of text in $\O(\log N + m)$ time and $O(z\log N)$ words of space. Perhaps 
it is possible to augment their data structure to also support rank and select. Also, is it
possible to improve their access time to $\O(\log N + m/\log_\sigma N)$ as we have done here
for grammar compressed strings?

\newpage
\bibliographystyle{splncs03}
\bibliography{grammar}

\newpage
\section{Appendix}
\label{sect:appendix}

\begin{figure}
\begin{center}
\includegraphics[width=0.8\linewidth]{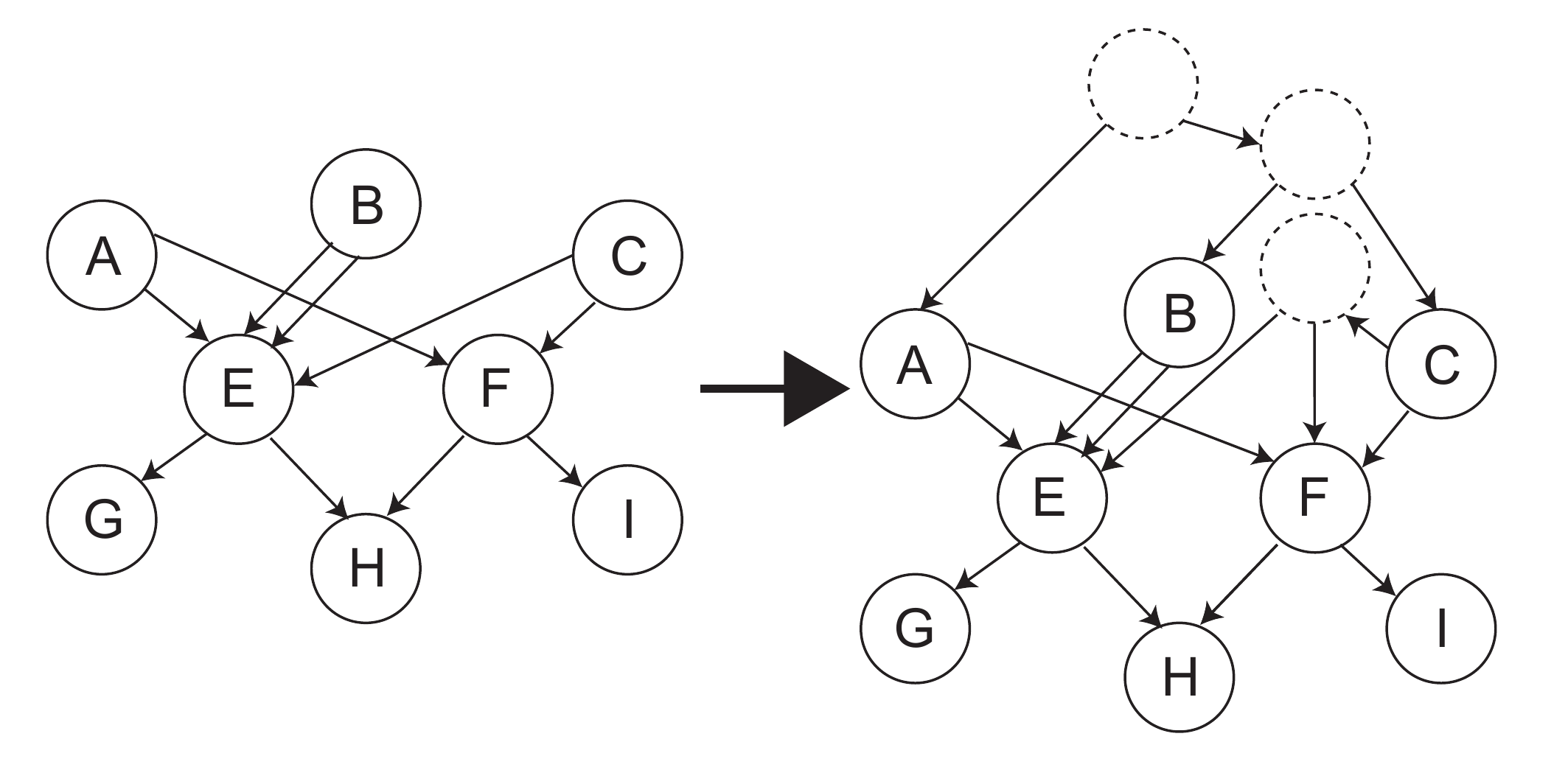}
\end{center}
\caption{An illustration of the transformation required in our reduction of $\rank$ queries
on a grammar compressed string to a path counting queries in DAGs.}
\label{fig-transform}
\end{figure}

\end{document}

%% file: heavypath.pdf_t
\begin{picture}(0,0)%
\includegraphics{heavypath.pdf}%
\end{picture}%
\setlength{\unitlength}{4144sp}%
\begingroup\makeatletter\ifx\SetFigFont\undefined%
\gdef\SetFigFont#1#2#3#4#5{%
  \reset@font\fontsize{#1}{#2pt}%
  \fontfamily{#3}\fontseries{#4}\fontshape{#5}%
  \selectfont}%
\fi\endgroup%
\begin{picture}(3624,3213)(3589,-3901)
\put(5041,-871){\makebox(0,0)[lb]{\smash{{\SetFigFont{12}{14.4}{\rmdefault}{\mddefault}{\updefault}{\color[rgb]{0,0,0}$P_0$}%
}}}}
\put(4051,-1321){\makebox(0,0)[lb]{\smash{{\SetFigFont{12}{14.4}{\rmdefault}{\mddefault}{\updefault}{\color[rgb]{0,0,0}$Q_1$}%
}}}}
\put(5851,-1321){\makebox(0,0)[lb]{\smash{{\SetFigFont{12}{14.4}{\rmdefault}{\mddefault}{\updefault}{\color[rgb]{0,0,0}$P_1$}%
}}}}
\put(5491,-1771){\makebox(0,0)[lb]{\smash{{\SetFigFont{12}{14.4}{\rmdefault}{\mddefault}{\updefault}{\color[rgb]{0,0,0}$P_2$}%
}}}}
\put(6481,-1771){\makebox(0,0)[lb]{\smash{{\SetFigFont{12}{14.4}{\rmdefault}{\mddefault}{\updefault}{\color[rgb]{0,0,0}$Q_2$}%
}}}}
\put(5401,-2671){\makebox(0,0)[lb]{\smash{{\SetFigFont{12}{14.4}{\rmdefault}{\mddefault}{\updefault}{\color[rgb]{0,0,0}$P_4$}%
}}}}
\put(5986,-2221){\makebox(0,0)[lb]{\smash{{\SetFigFont{12}{14.4}{\rmdefault}{\mddefault}{\updefault}{\color[rgb]{0,0,0}$Q_3$}%
}}}}
\put(5401,-2221){\makebox(0,0)[lb]{\smash{{\SetFigFont{12}{14.4}{\rmdefault}{\mddefault}{\updefault}{\color[rgb]{0,0,0}$P_3$}%
}}}}
\put(5626,-3121){\makebox(0,0)[lb]{\smash{{\SetFigFont{12}{14.4}{\rmdefault}{\mddefault}{\updefault}{\color[rgb]{0,0,0}$Q_5$}%
}}}}
\put(4906,-2671){\makebox(0,0)[lb]{\smash{{\SetFigFont{12}{14.4}{\rmdefault}{\mddefault}{\updefault}{\color[rgb]{0,0,0}$Q_4$}%
}}}}
\put(5401,-3121){\makebox(0,0)[lb]{\smash{{\SetFigFont{12}{14.4}{\rmdefault}{\mddefault}{\updefault}{\color[rgb]{0,0,0}$P_5$}%
}}}}
\end{picture}%

%% file: z.bbl
\begin{thebibliography}{10}
\providecommand{\url}[1]{\texttt{#1}}
\providecommand{\urlprefix}{URL }

\bibitem{BGIILL2012}
Bannai, H., Gagie, T., I, T., Inenaga, S., Landau, G.M., Lewenstein, M.: An
  efficient algorithm to test square-freeness of strings compressed by
  straight-line programs. Information Processing Letters  112(19),  711--714
  (2012)

\bibitem{bc2009}
Bendersky, M., Croft, W.B.: Finding text reuse on the web. In: Proc. 2nd
  International Conference on Web Search and Web Data Mining (WSDM). pp.
  262--271 (2009)

\bibitem{BCG2013}
Bille, P., Cording, P.H., G{\o}rtz, I.L.: Compact q-gram profiling of
  compressed strings. In: Proc. 24th Symposium on Combinatorial Pattern
  Matching (CPM). pp. 62--73. LNCS 7922 (2013)

\bibitem{BCG2014}
Bille, P., Cording, P.H., G{\o}rtz, I.L.: Compressed subsequence matching and
  packed tree coloring. In: Proc. 25th Symposium on Combinatorial Pattern
  Matching (CPM). pp. 40--49. LNCS 8486 (2014)

\bibitem{BLRSSW2011}
Bille, P., Landau, G.M., Raman, R., Sadakane, K., Satti, S.R., Weimann, O.:
  Random access to grammar-compressed strings. In: Proc. 22nd Symposium on
  Discrete Algorithms (SODA). pp. 373--389. SIAM (2011)

\bibitem{CLLPPSS2005}
Charikar, M., Lehman, E., Liu, D., Panigrahy, R., Prabhakaran, M., Sahai, A.,
  Shelat, A.: The smallest grammar problem. IEEE Transactions on Information
  Theory  51(7),  2554--2576 (2005)

\bibitem{CN2010}
Claude, F., Navarro, G.: Self-indexed grammar-based compression. Fundamenta
  Informaticae  111(3),  313--337 (2010)

\bibitem{CN2012}
Claude, F., Navarro, G.: Improved grammar-based compressed indexes. In: Proc.
  19th International Symposium on String Processing and Information Retrieval
  (SPIRE). pp. 180--192. LNCS 7608 (2012)

\bibitem{cohen2003reachability}
Cohen, E., Halperin, E., Kaplan, H., Zwick, U.: Reachability and distance
  queries via 2-hop labels. SIAM Journal on Computing  32(5),  1338--1355
  (2003)

\bibitem{1000genomes}
Consortium, .G.P.: A map of human genome variation from population-scale
  sequencing. Nature  467,  1061--1073 (2010)

\bibitem{FM2010}
Ferragina, P., Manzini, G.: On compressing the textual web. In: Proc. 3rd
  Conference on Web Search and Data Mining (WSDM). pp. 391--400 (2010)

\bibitem{FV2007}
Ferragina, P., Venturini, R.: A simple storage scheme for strings achieving
  entropy bounds. Theoretical Computer Science  372(1),  115--121 (2007)

\bibitem{FW90}
Fredman, M.L., Willard, D.E.: Blasting through the information theoretic
  barrier with fusion trees. In: Proceedings of the twenty-second annual ACM
  symposium on Theory of computing. pp. 1--7. ACM (1990)

\bibitem{FW93}
Fredman, M.L., Willard, D.E.: Surpassing the information theoretic bound with
  fusion trees. Journal of computer and system sciences  47(3),  424--436
  (1993)

\bibitem{GGKNP2012}
Gagie, T., Gawrychowski, P., K{\"a}rkk{\"a}inen, J., Nekrich, Y., Puglisi,
  S.J.: A faster grammar-based self-index. In: Proc. International Conference
  on Language and Automata Theory and Applications (LATA). pp. 240--251. LNCS
  7183 (2012)

\bibitem{GGP2011}
Gagie, T., Gawrychowski, P., Puglisi, S.J.: Faster approximate pattern matching
  in compressed repetitive texts. In: Proc. International Symposium on
  Algorithms and Computation (ISAAC). pp. 653--662. LNCS 7074 (2011)

\bibitem{G2012}
Gawrychowski, P.: Faster algorithm for computing edit distance between
  {SLP}-compressed strings. In: Proc. Symposium on String Processing and
  Information Retrieval (SPIRE). pp. 229--236. LNCS 7608 (2012)

\bibitem{GMR2006}
Golynski, A., Munro, J.I., Rao, S.S.: Rank/select operations on large
  alphabets: a tool for text indexing. In: Proc. 17th Symposium on Discrete
  Algorithms (SODA). pp. 368--373. SIAM (2006)

\bibitem{GBIT2013}
Goto, K., Bannai, H., Inenaga, S., Takeda, M.: Fast q-gram mining on {SLP}
  compressed strings. Journal of Discrete Algorithms  18,  89--99 (2013)

\bibitem{GGV2003}
Grossi, R., Gupta, A., Vitter, J.S.: High-order entropy-compressed text
  indexes. In: Proc. 14th Symposium on Discrete Algorithms (SODA). pp.
  841--850. SIAM (2003)

\bibitem{ggv2004}
Grossi, R., Gupta, A., Vitter, J.S.: When indexing equals compression:
  experiments with compressing suffix arrays and applications. In: Proc. 15th
  Symposium on Discrete Algorithms (SODA). pp. 636--645. SIAM (2004)

\bibitem{HLLW2013}
Hermelin, D., Landau, G.M., Landau, S., Weimann, O.: Unified compression-based
  acceleration of edit-distance computation. Algorithmica  65(2),  339--353
  (2013)

\bibitem{HPSTV2013}
Hon, W.K., Patil, M., Shah, R., Thankachan, S.V., Vitter, J.S.: Indexes for
  document retrieval with relevance. In: Space-Efficient Data Structures,
  Streams, and Algorithms. pp. 351--362. LNCS 8066 (2013)

\bibitem{IMSIBTNS2013}
I, T., Matsubara, W., Shimohira, K., Inenaga, S., Bannai, H., Takeda, M.,
  Narisawa, K., Shinohara, A.: Detecting regularities on grammar-compressed
  strings. In: Proc. 38th International Symposium on Mathematical Foundations
  of Computer Science (MFCS). pp. 571--582. LNCS 8087 (2013)

\bibitem{INIBT2013a}
I, T., Nakashima, Y., Inenaga, S., Bannai, H., Takeda, M.: Faster {L}yndon
  factorization algorithms for {SLP} and {LZ78} compressed text. In: Proc. 20th
  International Symposium on String Processing and Information Retrieval
  (SPIRE). pp. 174--185. LNCS 8214 (2013)

\bibitem{INIBT2013b}
I, T., Nishimoto, T., Inenaga, S., Bannai, H., Takeda, M.: Compressed automata
  for dictionary matching. In: Proc. 18th International Conference on
  Implementation and Application of Automata (CIAA). pp. 319--330. LNCS 7982
  (2013)

\bibitem{IB2012}
Inenaga, S., Bannai, H.: Finding characteristic substrings from compressed
  texts. International Journal of Foundations of Computer Science  23(2),
  261--280 (2012)

\bibitem{SLP}
Karpinski, M., Rytter, W., Shinohara, A.: An efficient pattern-matching
  algorithm for strings with short descriptions. Nordic Journal of Computing
  4,  172--186 (1997)

\bibitem{ml2000}
Larsson, N.J., Moffat, A.: Offline dictionary-based compression. Proc. of the
  IEEE  88(11),  1722--1732 (2000)

\bibitem{MNSV09jcb}
M{\"a}kinen, V., Navarro, G., Sir{\'e}n, J., V{\"a}lim{\"a}ki, N.: Storage and
  retrieval of highly repetitive sequence collections. Journal of Computational
  Biology  17(3),  281--308 (2010)

\bibitem{M2001}
Manzini, G.: An analysis of the {B}urrows-{W}heeler transform. Journal of the
  ACM  48(3),  407--430 (2001)

\bibitem{mbcmz2005}
Metzler, D., Bernstein, Y., Croft, W.B., Moffat, A., Zobel, J.: Similarity
  measures for tracking information flow. In: Proc. 14th ACM Int. Conf.
  Information and Knowledge Management (CIKM). pp. 517--524. ACM Press (2005)

\bibitem{M1996}
Munro, J.I.: Tables. In: Proc. 16th Conference on Foundations of Software
  Technology and Theoretical Computer Science (FSTTCS). pp. 37--42. LNCS 1180
  (1996)

\bibitem{n2012}
Navarro, G.: Indexing highly repetitive collections. In: Proc. 23rd
  International Workshop on Combinatorial Algorithms (IWOCA). pp. 274--279.
  LNCS 7643 (2012)

\bibitem{N2014b}
Navarro, G.: Spaces, trees and colors: The algorithmic landscape of document
  retrieval on sequences. ACM Computing Surveys  46(4),  article 52 (2014), 47
  pages

\bibitem{N2014a}
Navarro, G.: Wavelet trees for all. Journal of Discrete Algorithms  25,  2--20
  (2014)

\bibitem{NM2007}
Navarro, G., M{\"a}kinen, V.: Compressed full-text indexes. ACM Computing
  Surveys  39(1),  article 2 (2007)

\bibitem{Navarro14}
Navarro, G., Ord{\'o}{\~n}ez, A.: Grammar compressed sequences with rank/select
  support. In: Proceedings of the 21st International Symposium on String
  Processing and Information Retrieval (SPIRE) (2014), to appear

\bibitem{NPV2011}
Navarro, G., Puglisi, S.J., Valenzuela, D.: Practical compressed document
  retrieval. In: Proc. 10th International Symposium on Experimental Algorithms
  (SEA). pp. 193--205. LNCS 6630 (2011)

\bibitem{NPV2013}
Navarro, G., Puglisi, S.J., Valenzuela, D.: General document retrieval in
  compact space. ACM Journal of Experimental Algorithmics  (2014), to appear

\bibitem{NV2012}
Navarro, G., Valenzuela, D.: Space-efficient top-k document retrieval. In:
  Proc. 11th International Symposium on Experimental Algorithms (SEA). pp.
  307--319. LNCS 7276 (2012)

\bibitem{os2007}
Okanohara, D., Sadakane, K.: Practical entropy-compressed rank/select
  dictionary. In: Proc. 9th workshop on Algorithm Engineering and Experiments.
  pp. 60--70. SIAM (2007)

\bibitem{rrr2007}
Raman, R., Raman, V., Rao, S.S.: Succinct indexable dictionaries with
  applications to encoding {\it k}-ary trees, prefix sums and multisets. ACM
  Transactions on Algorithms  3(4) (2007)

\bibitem{rsh2004}
Riesenfeld, C., Schloss, P.D., , Handelsman, J.: Metagenomics: genomic analysis
  of microbial communities. Annual Review of Genetics  38,  525--552 (2004)

\bibitem{R2003}
Rytter, W.: Application of {Lempel-Ziv} factorization to the approximation of
  grammar-based compression. Theoretical Computer Science  302(1--3),  211--222
  (2003)

\bibitem{sg2006}
Sadakane, K., Grossi, R.: Squeezing succinct data structures into entropy
  bounds. In: Proc. 17th Symposium on Discrete Algorithms (SODA). pp.
  1230--1239. SIAM (2006)

\bibitem{T2011}
Tiskin, A.: Towards approximate matching in compressed strings: Local
  subsequence recognition. In: Proc. 6th International Computer Science
  Symposium in Russia (CSR). pp. 401--414. LNCS 6651 (2011)

\bibitem{VY2013}
Verbin, E., Yu, W.: Data structure lower bounds on random access to
  grammar-compressed strings. In: Proc. 24th Symposium on Combinatorial Pattern
  Matching (CPM). pp. 247--258. LNCS 7922 (2013)

\bibitem{ZL77}
Ziv, J., Lempel, A.: A universal algorithm for sequential data compression.
  IEEE Transactions on Information Theory  23(3),  337--343 (1977)

\end{thebibliography}
